\documentclass[aps, prb, twocolumn, superscriptaddress, floatfix, groupedaddress]{revtex4-2} % showkeys, showpacs

\usepackage{graphicx}
\graphicspath{{figures/}}
\usepackage[utf8]{inputenc}
\usepackage{amssymb, amsmath, commath, braket, physics, soul}
\usepackage{hyperref}
\hypersetup{
colorlinks=true,
linkcolor=red
,
filecolor=blue,
urlcolor=magenta,
}
\usepackage{placeins, standalone}
\usepackage{soul}
\usepackage{xcolor}
\definecolor{darkblue}{rgb}{0,0,.65}
\definecolor{darkgreen}{rgb}{0.3,0.6,0.3}
\definecolor{darkorange}{rgb}{0.85,0.65,0.3}
\definecolor{cyan1}{rgb}{0.0, 0.6, 0.6}

\newcommand{\AKD}[1]{{\color{blue}(AKD: #1)}}

\newcommand{\bte}{$\beta$-ensemble}

\newcommand{\corrvoid}{correlation void}

\newcommand{\Eth}{\ensuremath{E_{\textrm{Th}}}}	%	Thouless energy
\newcommand{\Ecal}{\ensuremath{\mathcal{E}}}	%	Unfolded energy
\newcommand{\Egap}{\ensuremath{\Delta\Ecal}}	%	Unfolded energy gap
\def\eg{\ensuremath E_{\mathrm{G}}}				%	Griffiths energy

\newcommand{\fbslj}[2]{\ensuremath{\mathrm{J}_{#1}\del{#2}}}

\newcommand{\fgamma}[1]{\ensuremath{\Gamma\del{#1}}}

\newcommand\gEM{ \ensuremath{\gamma_\mathrm{E}} }

\newcommand\iiserk{Department of Physical Sciences, Indian Institute of Science Education and Research Kolkata, Mohanpur 741246, India}

\newcommand\mean[1]{\ensuremath{\left\langle #1 \right\rangle}}

\newcommand{\prob}[1]{\ensuremath{\textrm{P}\del{#1}}}

\newcommand{\sff}[1]{\ensuremath{\mathcal{K}\del{#1}}}
\newcommand\tTh{\ensuremath{t_\mathrm{Th}}}
\newcommand\tdip{\ensuremath{t_\mathrm{dip}}}
\newcommand\tH{\ensuremath{t_\mathrm{H}}}
\newcommand\tR{\ensuremath{t_\mathrm{R}}}

\newcommand\tauH{\ensuremath{\tau_\mathrm{H}}}
\newcommand\tauR{\ensuremath{\tau_\mathrm{R}}}

\begin{document}

\title{Anomalous energy correlations and spectral form factor in the nonergodic phase of the \bte}
\author{Basudha Roy}\email{br23rs008@iiserkol.ac.in}
\author{Adway Kumar Das}\email{akd19rs062@iiserkol.ac.in}
\author{Anandamohan Ghosh}\email{anandamohan@iiserkol.ac.in}
\affiliation{\iiserk}
\author{Ivan M. Khaymovich}\email{ivan.khaymovich@gmail.com}
\affiliation{Nordita, Stockholm University and KTH Royal Institute of Technology Hannes Alfv\'ens v\"ag 12, Sx-106 91 Stockholm, Sweden}
%\affiliation{Institute for Physics of Microstructures, Russian Academy of Sciences, 603950 Nizhny Novgorod, GSP-105, Russia}

%=================================
\begin{abstract}
	The \bte\ is a prototypical model of a single particle system on a one-dimensional disordered lattice with inhomogeneous nearest neighbor hopping. 
	% The \bte\ exhibits two second order phase transitions: ergodicity breaks down at $\gamma = 0$ and delocalization-localization transition at $\gamma = 1$ where the Dyson index $\beta = N^{-\gamma}$ and $N$ is the system size. 
	Corresponding nonergodic phase has an anomalous critical energy scale, $E_c$: correlations are present above and absent below $E_c$ as reflected in the number variance. We study the dynamical properties of the \bte\ where the critical energy controls the characteristic timescales. In particular, the spectral form factor equilibrates at a relaxation time, $\tR \equiv E_c^{-1}$, which is parametrically smaller than the Heisenberg time, $\tH$, given by the inverse of the mean level spacing. Incidentally, the dimensionless relaxation time, $\tauR\equiv \tR/\tH \ll 1$ is equal to the Dyson index, $\beta$. We show that the energy correlations are absent within a temporal window $\tR < t < \tH$, which we term as the {\itshape\corrvoid}. This is in contrast to the mechanism of equilibration in a typical many-body system. We analytically explain the qualitative behavior of the number variance and the spectral form factor of the \bte\ by a spatially local mapping to the Anderson model.
\end{abstract}

%============================================================
% 	PACS, the Physics and Astronomy Classification Scheme.
% 	05.45.Mt: Quantum chaos; semiclassical methods
%	02.10.Yn: Matrix Theory
%	89.75.Da: Systems obeying scaling laws
%--------------------------------------------------------
%	Use showkeys class option if keyword display desired

\maketitle
%=================================
\section{Introduction}
%=================================
In an isolated quantum mechanical system, the non-equilibrium dynamics of an observable is dictated by the energy correlations and the nature of hybridization among the energy eigenstates. 
%, their localization lengths. 
Particularly, in ergodic systems~\cite{Neumann2010, Goldstein2010}, Haar distributed bulk energy eigenstates exhibit strong mutual hybridization, leading to diffusive dynamics, thermalization of local observables~\cite{Srednicki1994, Rigol2008, DAlessio2016, Deutsch2018} and quantum chaos via correlated energy levels~\cite{Bohigas1984, Borgonovi2016}. 

In nonergodic systems, loss of ergodicity leads to a rich set of behaviors: (multi)fractal scaling of the eigenstate fluctuations~\cite{Evers2008, Mace2019, Solorzano2021, BastarracheaMagnani2024, Cugliandolo2024, Das2025a, Das2022b}, violation of thermalization~\cite{Mori2018, Vidmar2016, Sinha2020, Dong2023}, emergent integrals of motion~\cite{Serbyn2013, Modak2016, Kulshreshtha2018, Abanin2019, Shtanko2025}, % Prosen2013
anomalous diffusion~\cite{Santos2016, Rosenberg2024,Bar2015,Luitz2016}, heterogeneous energy correlations yielding critical energy scales~\cite{Soukoulis1984, Erdoes2015, Jagannathan2023}. Specifically, the long-range energy correlations manifest as a dip in the time evolution of an observable below its equilibrium value, known as the {\itshape correlation hole}~\cite{Leviandier1986, Pique1987, Alhassid1992, Michaille1999, TorresHerrera2017, Schiulaz2019, Das2023a}. Such non-equilibrium features are accessible in various experimental setups, e.g.~ion trap~\cite{Smith2016}, superconducting qubit~\cite{Dong2025}, molecular spectroscopy~\cite{Leviandier1986}, elucidating the role of energy correlations in strongly correlated many-body systems~\cite{Das2025, VallejoFabila2024, VallejoFabila2025arxiv, ZarateHerrada2025arxiv}.

A random matrix model with an explicit control over the energy correlations is the \bte, with a joint density of the energy levels~\cite{Dumitriu2002,Forrester2010book, Mehta2004book}
\begin{align}
	\begin{split}
		\prob{E_1, E_2, \dots, E_N} &= \frac{1}{\mathcal{Z}_\beta} e^{-\sum\limits_{i=1}^{N} \frac{E_i^2}{2} + \beta \sum\limits_{i<j} \log|E_i - E_j|} %\prod_{i<j}\abs{E_i - E_j}^\beta
		%\mathcal{Z}_\beta &= (2\pi)^\frac{N}{2}\prod_{j = 1}^{N}\frac{ \fgamma{1+\frac{\beta}{2}j} }{ \fgamma{1 + \frac{\beta}{2}} }
	\end{split}
	\label{eq:beta_jpdf}
\end{align}
where $\mathcal{Z}_\beta$ is the normalization constant and the Dyson's index $\beta$, dictates the energy correlation via pairwise logarithmic repulsions. Eq.~\eqref{eq:beta_jpdf} describes the energy levels of \bte, a disordered single-particle system on a one-dimensional (1D) lattice, which exhibits two second order phase transitions despite having uncorrelated short-range hoppings. 
Moreover, an extended parameter regime exhibits nonergodicity along with non-trivial long-range energy correlations~\cite{Das2022, Das2023, Das2024, Das2025a}. The dynamical signatures of such energy correlations and the associated timescales %in the nonergodic phase of the \bte\ 
are the focus of this paper.

%=================================
%	Schematic for energy correlation -> SFF
\begin{figure*}[t]
	\centering
    \includegraphics[width=0.85\textwidth]{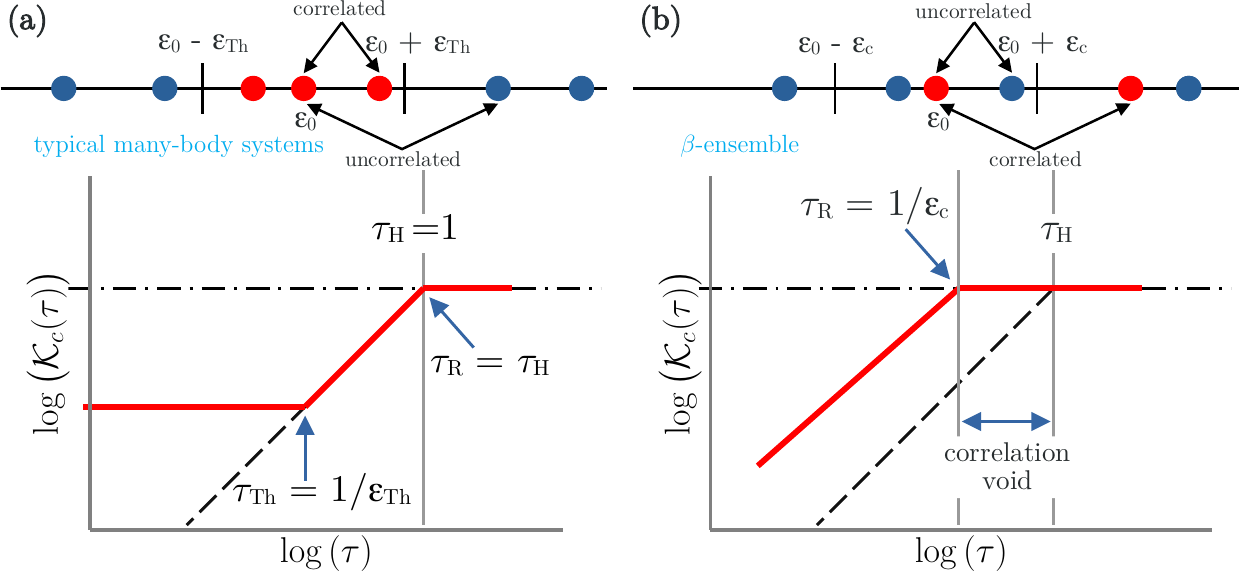}
	\caption{Schematics of energy correlations governing spectral form factor (SFF). Dimensionless time $\tau$ is Fourier dual to the unfolded energy level $\Ecal$. (Top) Red energy levels are mutually correlated but uncorrelated to any blue energy level.(Bottom) Red bold curve shows the time evolution of the connected SFF, $\mathcal{K}_c$ % (Eq.~\eqref{eq:SFF_parts})
    in log-log scale where the straight dashed (horizontal dot-dashed) line corresponds to GOE (Poisson value). (a)~Typical many-body system: % (Eq.~\eqref{eq:nvar_typ}).
    given an energy $\Ecal_0$, any energy level within $[\Ecal_0 - \Ecal_\mathrm{Th}, \Ecal_0 + \Ecal_\mathrm{Th}]$ is correlated to $\Ecal_0$ and uncorrelated otherwise. 
		%========
		(b)~\bte\ in the nonergodic phase: %(Eq.~\eqref{eq:nvar_beta}). 
        any energy level within $[\Ecal_0 - \Ecal_c, \Ecal_0 + \Ecal_c]$ is necessarily uncorrelated to $\Ecal_0$ while correlated levels exist outside the energy window. Corresponding SFF equilibrates at $\tauR$. The interval $[\tauR, \tauH]$ is denoted as the \corrvoid.
	}
	\label{fig:Schematic_SFF}
\end{figure*}
%=================================

%	Synopsis
In a nonergodic many-body system, the interplay of disorder and interaction may yield critical energy like the mobility edge~\cite{Soukoulis1984, Luitz2015}, or critical energy scales, e.g.~Thouless~\cite{Erdoes2015, Sierant2020, Das2022a, Corps2020}, Altshuler-Shklovskii energy scale~\cite{Sivan1987, Aronov1995, Jagannathan2023}. In particular, any two energy levels with gap smaller than the Thouless energy, $\Eth$, are correlated as in the random matrix models and uncorrelated otherwise. Since energy and time are Fourier dual to each other, the presence of Thouless energy scale is reflected in the dynamics of a system. For example, dynamical measures of energy correlations, e.g.~the spectral form factor (SFF), exhibit a universal behavior predicted by the random matrix theory between the Thouless time, $\tTh$ (inverse of the Thouless energy) and the Heisenberg time, $\tH$, (inverse of the mean level spacing) while equilibrating beyond $\tH$ in case of typical many-body systems. 
%In contrast, we intend to study the long-range correlations in the energy spectrum of \bte\ in order to identify a critical energy scale complementary to the Thouless energy—i.e., an energy scale that separates the uncorrelated and correlated spectral regimes. Correspondingly, we investigate the SFF to determine the associated characteristic timescale that governs the dynamical transition between these regimes.
In contrast, the nonergodic phase of the \bte\ exhibit a multi-scale energy spectrum with a critical energy scale, $E_c$, behaving as a counterpart to the Thouless energy. The partial long-range correlations present above $E_c$ leads to a correlation hole in the SFF but the absence of correlation below $E_c$ leads to an equilibration long before $\tH$, producing a {\itshape \corrvoid}, see Fig.~\ref{fig:Schematic_SFF}. Thus, the relaxation mechanism of the \bte\ is different than that of a typical many-body system~\cite{Schiulaz2019, Sierant2020, Hopjan2023} or random matrix models like the Rosenzweig-Porter ensemble~\cite{Rosenzweig1960, Kravtsov2015, Monthus2017, Venturelli2023,RP_R(t)_2018,Amini2017}. 

In this work, we look at the SFF as a dynamical quantity reflecting the energy correlations across all possible length scales~\cite{Mehta2004book, Guhr1998}, hence, an efficient probe of quantum chaos~\cite{Dag2023, Das2025, Fritzsch2025Arxiv, Kalsi2025, Vikram2023}. Such a dynamical measure has also been used to probe the stability of the many-body localized phase~\cite{Suntajs2020}, breaking of ergodicity~\cite{Bertrand2016, Sierant2019, Sierant2020, Das2025b}. %, and scale-invariant dynamics~\cite{Hopjan2023, Hopjan2023a}. 
The SFF is experimentally measured in atomic and molecular spectroscopy~\cite{Leviandier1986, Guhr1990, Michaille1999}, Floquet superconducting circuit~\cite{Dong2025}, nonlinear optical platform~\cite{Wang2025}. In a generic many-body setup, SFF can be measured via non-demolition measurement of ancilla qubit~\cite{Vasilyev2020} and randomized measurements~\cite{Joshi2022}. We will show that the correlation hole of the SFF serves as a dynamical quantity to identify the delocalization-localization transition in case of the \bte.

The organization of the paper is as follows: in Sec.~\ref{sec:model}, we introduce the \bte\ and discuss its spatially local equivalence to the 1D Anderson model to explain the phase transitions. In Sec.~\ref{sec:energy_corr}, we show the multi-scale structure of the energy spectrum leading to the critical energy scale, $E_c$. In Sec.~\ref{sec:num_var}, we demonstrate how $E_c$ dictates the long-range energy correlations while behaving complimentary to the Thouless energy. In Sec.~\ref{sec:SFF}, we discuss the SFF, its non-equilibrium features like the power-law decays, correlation hole and characteristic time scales. Importantly, we show that the relaxation time of the SFF is dictated by the inverse of $E_c$ obtained from the spectral statistics. 
% Consequently, the mechanisms of relaxation of a non-equilibrium dynamics is different for the \bte\ from a typical many-body system, as explained in Fig.~\ref{fig:Schematic_SFF}. 
Our concluding remarks are given in Sec.~\ref{sec:discuss}.

%=================================
\section{\bte}\label{sec:model}
\begin{figure}[t]
	\centering
	\includegraphics[width=\columnwidth, clip=true, trim=1.5 1.5 1 1.5]{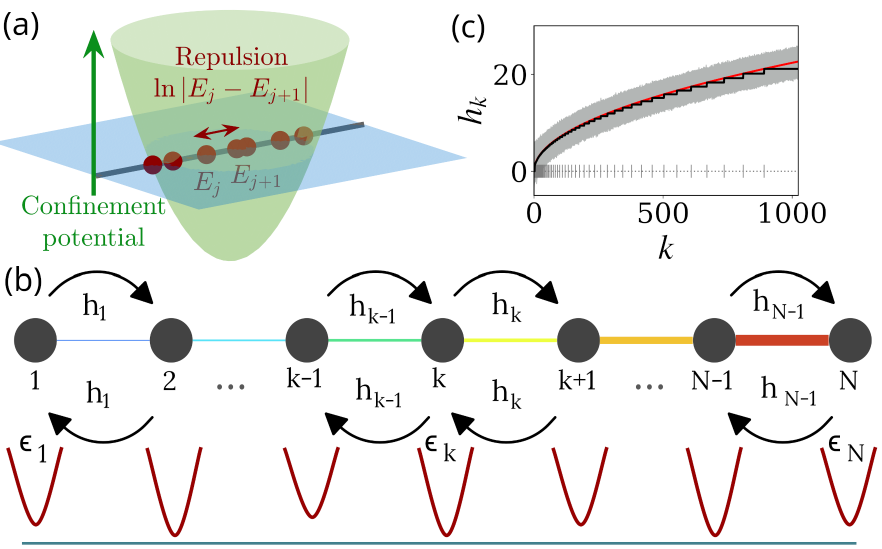}
	\caption{(a)~2D Coulomb gas confined on a line under harmonic potential with equilibrium configuration given by Eq.~\eqref{eq:beta_jpdf}.
		%--------------------------
		(b)~1D lattice corresponding to the Hamiltonian in Eq.~\eqref{eq:H_beta}. Disordered on-site potential is homogeneous while the nearest-neighbor hopping terms increase along the lattice (thickness, color of links).
		%--------------------------
		(c)~Shaded region shows 95\% confidence interval of the random hopping terms with a typical behavior shown via red line. Black step function shows the local approximation of the hopping terms, leading to the emergence of spatial blocks as in Eq.~\eqref{eq:block}, shown via vertical bars. 
	}
	\label{fig:lattice}
\end{figure}

%=================================
The energy levels of the \bte\ correspond to the equilibrium configuration of a classical 2D Coulomb gas confined on a line at a temperature $\beta^{-1}$~\cite{Dyson1962a}, as shown in Fig.~\ref{fig:lattice}(a). %experiencing mutual logarithmic repulsion and a harmonic confining potential~\cite{Dyson1962a}. 
Such a static lattice gas model has been widely studied~\cite{Forrester2009book, Forrester2010book, Forrester2011book} with a focus on the extreme eigenvalues~\cite{Borot2011, Forrester2012b, Allez2014, Edelman2014}, connection to stochastic differential operators~\cite{Edelman2007, Bourgade2014a, Bekerman2015, Krishnapur2016} and the density of states~\cite{Desrosiers2006, Forrester2012, Allez2012, Witte2014, Forrester2017}. The Coulomb gas analogy puts the normalization constant in Eq.~\eqref{eq:beta_jpdf} as a canonical partition function~\cite{Livan2018book, Das2019} such that starting from a random configuration of Coulomb gas particles and minimizing the associated free energy via Monte-Carlo simulations, one can obtain the energy levels of the \bte~\cite{Chafai2019, Pandey2017}.

%==============================
%	Matrix model
Dumitriu and Edelman proposed an ensemble of real symmetric random tridiagonal matrices, $\cbr{H_\beta}$ with the same joint density of energy levels as in Eq.~\eqref{eq:beta_jpdf}~\cite{Dumitriu2002}. For $\beta = 1, 2$ and 4, $H_\beta$ is the Krylov space representation of a Wigner-Dyson matrix from the orthogonal, unitary and symplectic symmetry classes, respectively~\cite{Balasubramanian2023, Parker2019}. The tridiagonal matrix $H_\beta$ can also be understood as the single-particle representation of the Hamiltonian% in the site basis, $\cbr{\ket{n}}$
\begin{align}
	\hat{H} = \sum_{k = 1}^{N} \epsilon_k \hat{n}_k + \sum_{k = 1}^{N-1} h_k \del{ \hat{c}_k^\dagger \hat{c}_{k+1} + \hat{c}_k \hat{c}_{k+1}^\dagger }%\sum_{n = 1}^{N} \epsilon_n \ket{n}\bra{n} + \sum_{n = 1}^{N-1} h_n \del{ \ket{n}\bra{n+1} + \mathrm{H.c.} }
	\label{eq:H_beta}
\end{align}
where $\hat{c}_k^\dagger$ ($\hat{c}_k$) is the creation (annihilation) operator and $\hat{n}_k \equiv \hat{c}_k^\dagger\hat{c}_k$ is the number operator acting on the $k$th site of a disordered 1D lattice with open boundary, see Fig.~\ref{fig:lattice}(b). The random on-site potential, $\epsilon_k$ follows the normal distribution such that $\mean{\epsilon_k^2} = 1$. The random nearest-neighbor hopping strength $h_k$ is such that $\sqrt{2} h_k$ follows the Chi-distribution with $k\beta$ degrees of freedom and increases on average with $k$. Such an inhomogeneous 1D lattice can be experimentally realized in superconducting processors with individual microwave controls~\cite{Dong2023}, cold atoms in tunable trap~\cite{Beugnon2007, Kaufman2021}, coupled oscillators~\cite{Dyson1953}, synthetic optical waveguides~\cite{Gao2023, Gao2025}.

%===============================================
%	Spatially local mapping to 1D Anderson model

In a typical 1D disordered system with uncorrelated short-range hopping, the energy states are exponentially localized with uncorrelated energy levels~\cite{Anderson1958} while phase transition is forbidden~\cite{Cuesta2004}. However, the inhomogeneity in the \bte\ yields second order phase transitions in the typical bulk properties: $\gamma = 1$ (delocalization-localization transition) and $\gamma = 0$ (ergodic-nonergodic transition) where $\beta = N^{-\gamma}$~\cite{Das2022}.

The energy levels of the single-particle model in Eq.~\eqref{eq:H_beta} follows the same joint density of energy levels as in Eq.~\eqref{eq:beta_jpdf}, where the energy variance is $(1+N^{1-\gamma})/2$~\cite{Dumitriu2002}. 
In order to make the global mean level spacing unity, we scale the Hamiltonian from the \bte\ with the following factor
\begin{align}
    \begin{split}
        \sigma_{\gamma} &= \frac{N}{2\sqrt{4 + 2N^{1-\gamma}}}\\ 
        \Rightarrow \sigma_\gamma &\sim N^{\frac{1+\gamma}{2}} \textrm{ for } 0 < \gamma < 1.
    \end{split}
    \label{eq:scale_H}
\end{align}
%The on-site potential at the $k$th site is rescaled to $\sigma_{\gamma} \epsilon_k$ 
Then, all the on-site potentials $\sim\mathcal{O}(\sqrt{N^{1+\gamma}})$ while the hopping amplitudes typically behave as~\cite{Das2023}
\begin{align}
	\mean{\abs{h_k}} \sim \begin{cases}
		\sqrt{N^{1+\gamma}}\exp\del{-N^{\gamma-1}}, & k < N^\gamma\\
		\sqrt{k N}, & k> N^\gamma
	\end{cases}.
    \label{eq:hop_typical}
\end{align}
Above equation implies that for the lattice indices within $[k, k+\delta k]$, the hopping amplitudes are approximately constant provided $\delta k \ll k$ and $\gamma>0$, as shown in Fig.~\ref{fig:lattice}(c). Then, the lattice governed by $\hat{H}$ can be split into nearly independent spatial blocks $\Delta_0, \Delta_1,\dots, \Delta_{l_\mathrm{max}}$ where $\Delta_0 = [1, N^\gamma]$ and $1\leq l \leq l_\mathrm{max} \equiv \del{1-\gamma}\ln N/ \ln c$
\begin{align}
	\Delta_l \equiv [N^{\gamma + \zeta_l}, N^{\gamma + \zeta_{l+1}}],\quad \zeta_l = (l-1) \ln c/ \ln N
	\label{eq:block}
\end{align}
given a constant $c\sim \mathcal{O}(1)$~\cite{Das2023}. Within the first spatial block $\Delta_0$ hopping terms are negligibly small compared to $\epsilon_k^\mathrm{typ} \sim \mathcal{O}(\sqrt{N^{1+\gamma}})$) such that the respective lattice sites are decoupled and spanned by $\delta$-localized eigenstates. Corresponding energy levels are uncorrelated and approximately equal to the on-site potentials. The energy levels coming from $\Delta_0$ lie within an energy window of size $\mathcal{O}(\sqrt{N^{1+\gamma}})$ %$\mathcal{O}(\sigma_{\gamma})$ 
centered around the zero energy. 

%=================================
%	Lattice
%=================================

\begin{figure}[t]
	\centering
	\includegraphics[width=\columnwidth]{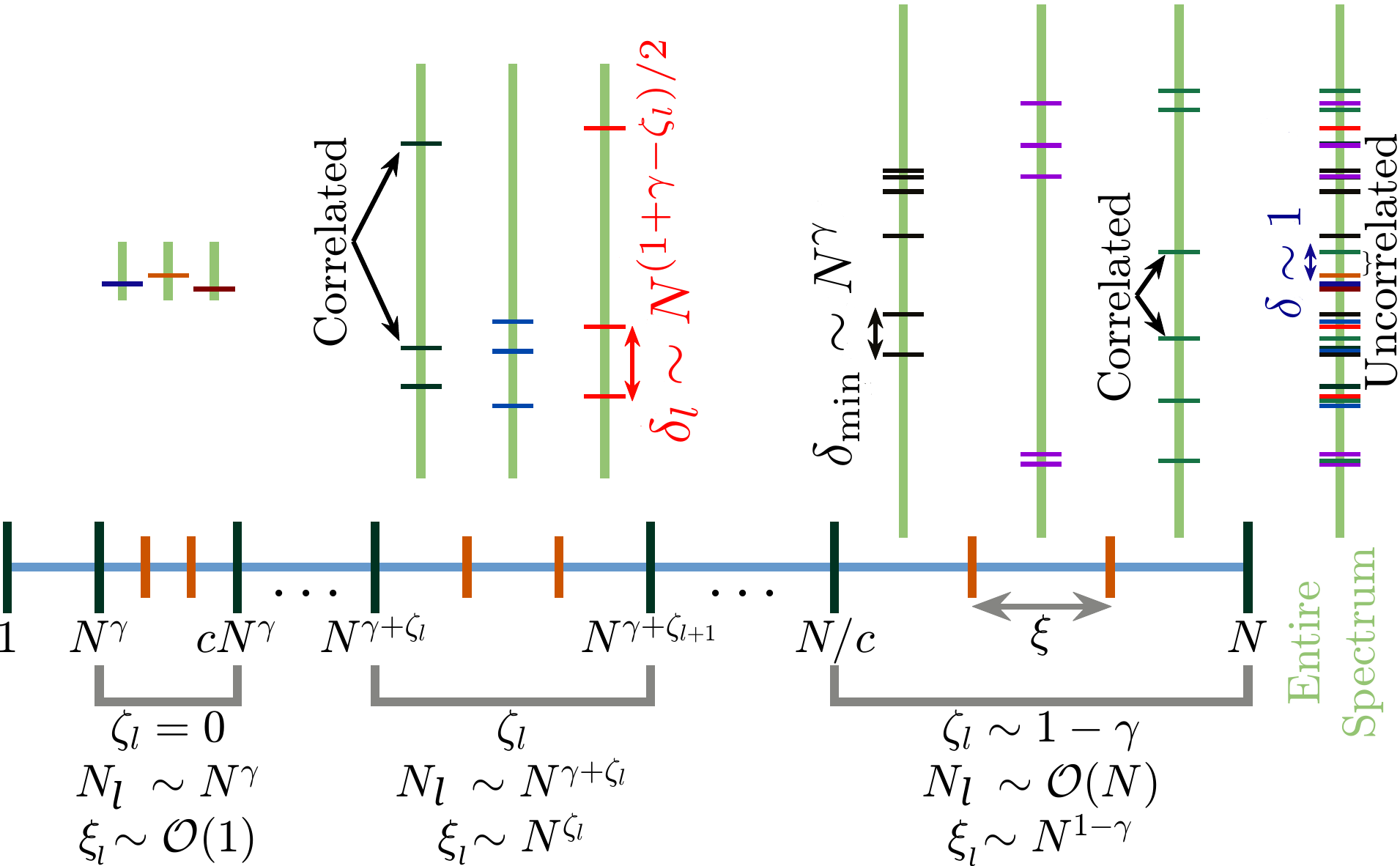}
	\caption{Spatially resolved spectrum structure. Emergent blocks (blue partitions on the light blue horizontal line) and sub-blocks (orange partitions) in a 1D lattice governed by the \bte\ in the NEE phase ($0 < \gamma < 1$). $N_{l}$ is the number of sites in the $l$th block, $\Delta_l$, given in Eq.~\eqref{eq:block}. Typical bulk localization length $\xi_l$ of the block $\Delta_l$ determines the number of sites in the respective sub-blocks as $N_l/\xi_l$. Corresponding local spectrum in each sub-block (ticks on the corresponding vertical dark green line) contains correlated energies, shown via same color. In the global spectrum with mean level spacing $\delta$, any two energy levels with different colors are uncorrelated.}
	\label{fig:partition}
\end{figure}

%On the other hand, 
For $l\geq 1$, $\Delta_l$ approximately describes a 1D Anderson model on $N^{\gamma+\zeta_l}$ number of sites with uncorrelated diagonal disorder and homogeneous hopping strength $N^{(1+\gamma+\zeta_l)/2}$. 
%disorder $\sim\mathcal{O}(1)$ in the on-site potentials. 
The energy states with support over $\Delta_l$ decay exponentially with a localization length~\cite{Izrailev1998}
\begin{align}
	\xi_l \sim  \left(\frac{\Gamma(1/4)}{\Gamma(3/4)}\right)^2\frac{\mean{h_k^2}}{\mean{\epsilon_k^2}} \simeq 8.75\frac{\mean{h_k^2}}{\mean{\epsilon_k^2}}
    %\frac{105 \mean{h_k^2}}{12 \mean{\epsilon_k^2}} 
    \sim N^{\zeta_l}.
	\label{eq:xi_sub_block}
\end{align}
Thus, the block $\Delta_l$ can be further separated into $N^\gamma$ sub-blocks of length $N^{\zeta_l}$. The energy states with localization centers within each sub-block hybridize with each other, leading to a density of states (DOS) of bandwidth
\begin{align}
    \sigma_l \simeq \sqrt{\frac{1}{N_l}\sum_{m\in\Delta_l} |H_{mn}|^2} \sim N^{\frac{1+\gamma+\zeta_l}{2}}.
\end{align}
Then, the mean level spacing of the $l$th spatial block is~\cite{Das2023}
\begin{align}
    \delta_l \sim \frac{N^{\frac{1+\gamma+\zeta_l}{2}}}{N^{\zeta_l}} = N^{\frac{1+\gamma-\zeta_l}{2}}.
    \label{eq:delta_block}
\end{align}
The emergent blocks and sub-blocks in a 1D lattice from the spatially local mapping of the \bte\ to the 1D Anderson model are schematically shown in Fig.~\ref{fig:partition}.

For an eigenstate $\ket{\Psi}$, the $q$th moment of the intensities scales as 
\begin{align}
    \sum_n |\Psi(n)|^{2q} \sim N^{D_q(1-q)}
\end{align}
where $\Psi(n) \equiv \braket{n}{\Psi}$ is the $n$th component in the basis $\cbr{\ket{n}}$ and $D_q$ is the $q$th fractal dimension. Such scaling exponents can be analytically estimated from the spatially local mapping of the \bte\ to the Anderson model for $0 < \gamma < 1$. Eqs.~\eqref{eq:block} and \eqref{eq:xi_sub_block} imply that the spatially largest block, $\Delta_{l_\mathrm{max}}$ has the maximal energy bandwidth $\sigma_{l_{\mathrm{max}}} \sim N$, hosts a finite fraction $\mathcal{O}(N)$ of all the states with localization length $N^{1-\gamma}$ and thus, dictates the typical behavior of the \bte. Then, 
%the DOS of the \bte\ has a width \BR{$\sim \frac{1}{\sigma_{\gamma}}N^{\frac{1-\gamma}{2}}$} while 
a typical bulk energy state of the \bte\ decays exponentially over a localization length $N^{1-\gamma}$. Such energy states are extended over an extensive part but vanishing fraction of all sites (i.e.~nonergodic). Thus, the regime $0<\gamma<1$ is the nonergodic extended (NEE) phase of the \bte, where a typical bulk energy state is fractal with a dimension $D_q^\mathrm{typ} = 1-\gamma$ for all values of $q>0$~\cite{Das2022}. 

In the next section, we look at the nature of energy correlations across all length scale to understand the spatially resolved energy spectrum shown in Fig.~\ref{fig:partition}.

%=================================

%=================================
\section{Energy correlation}\label{sec:energy_corr}
%=================================
In a spectrum with $N$ energy levels, $\cbr{E_n}$, the DOS is defined as $\rho\del{E} = N^{-1} \sum_{n = 1}^{N} \delta\del{E - E_n}$. For \bte, the ensemble averaged DOS is~\cite{Nakano2018, Allez2012}
\begin{align}
	\begin{split}
		\mean{\rho(E)} &= \frac{\sqrt{2}}{ \pi \alpha \sigma_{\gamma}} \frac{\exp\del{-\frac{\tilde{E}^2}{2}}}{ f(\alpha, \tilde{E}) },\quad \alpha \equiv \frac{N\beta}{2}, \quad \tilde{E}\equiv\frac{E}{\sigma_{\gamma}}\\
		f(\alpha, \tilde{E}) &= \frac{\fgamma{\frac{\alpha}{2}}}{\fgamma{\frac{1+\alpha}{2}}} {}_1F_{1}\del{\frac{\alpha}{2}, \frac{1}{2}; -\frac{\tilde{E}^2}{2} }^2 \\&+ 2 \tilde{E}^2 \frac{\fgamma{\frac{1+\alpha}{2}}}{\fgamma{\frac{\alpha}{2}}} {}_1F_{1}\del{\frac{1+\alpha}{2}, \frac{3}{2}; -\frac{\tilde{E}^2}{2} }^2
		%\\\rho(E) &= \frac{2^{2-\alpha}\fgamma{\alpha}}{ \sqrt{2\pi} \alpha } \frac{\exp\del{-\frac{E^2}{2}}}{ f(\alpha, E) },\quad \alpha \equiv \frac{N\beta}{2}\\
		%f(\alpha, E) &= \fgamma{\frac{\alpha}{2}}^2 {}_1F_{1}\del{\frac{\alpha}{2}, \frac{1}{2}, -\frac{E^2}{2} }^2 \\&+ 2 E^2\fgamma{\frac{1+\alpha}{2}}^2 {}_1F_{1}\del{\frac{1+\alpha}{2}, \frac{3}{2}, -\frac{E^2}{2} }^2\\
		% &\sqrt{\frac{\alpha}{\Gamma(\alpha)}} \int_{0}^{\infty} dt\: t^{\alpha - 1} \exp\del{-\frac{t^2}{2} + iE t}
	\end{split}
	\label{eq:DOS_beta}
\end{align}
where ${}_1F_{1}$ is Kummer confluent hypergeometric function. 
%For $\alpha\gg 1$, $\frac{\fgamma{\frac{1+\alpha}{2}}}{\fgamma{\frac{\alpha}{2}}}\approx \sqrt{\frac{\alpha}{2}}$. 
Eq.~\eqref{eq:DOS_beta} converges to the Wigner's semicircle law (normal distribution) for $\alpha \equiv {N\beta}/{2}\to \infty$ ($\alpha\to 0$).
% , while a system size invariant DOS exists at the localization transition point, $\gamma = 1$ (i.e.~$\beta = N^{-1}$)
% \begin{align}
% 	\begin{split}
% 		\mean{\rho(E)} &= \frac{2\sqrt{2}}{ \abs{\fbslk{\frac{1}{4}}{-\frac{\tilde{E}^2}{2}}^2 \tilde{E} } } = \frac{\frac{4\sqrt{2}}{\pi^2 |\tilde{E}|}}{\fbsli{-\frac{1}{4}}{\frac{\tilde{E}^2}{2}}^2 + \fbsli{\frac{1}{4}}{\frac{\tilde{E}^2}{2}}^2}
% 	\end{split}
% 	\label{eq:DOS_g_1}
% \end{align}}
% where $\fbsli{\theta}{x}$ ($\fbslk{\theta}{x}$) is the modified Bessel function of the 1st (2nd) kind of order $\theta$. 
In the thermodynamic limit ($N\to\infty$), the DOS of \bte\ follows semicircle (Gaussian) distribution for $\gamma < 1$ ($\gamma > 1$). The tridiagonal structure of a matrix from the \bte\ implies that the DOS can be obtained using Sturm sequence without doing exact diagonalization~\cite{Albrecht2009}. The standard deviation of energy w.r.t.~the DOS in Eq.~\eqref{eq:DOS_beta} is~\cite{Das2023}

\begin{align}
	\sqrt{\mean{E^2} - \mean{E}^2} = \frac{N}{4},\quad \mean{E^k} \equiv \int dE\; E^k\rho(E).
	\label{eq:E_var}
\end{align}

In the NEE phase ($0<\gamma<1$) of the \bte, the local mean level spacings of the spatial blocks, $\Delta_l$ (Eq.~\eqref{eq:delta_block}) decrease with $l$ with the smallest value $\delta_\mathrm{min} = N^\gamma$ present in $\Delta_{l_\mathrm{max}}$. However, the global mean level spacing, $\delta \sim 1$ 
% \frac{1}{\sigma_{\gamma}} N^{-\frac{1+\gamma}{2}}
as the energy spectrum of the \bte\ is the superposition of the local spectra from all possible sub-blocks. As $\delta < \delta_\mathrm{min}$, neighboring energy levels in the global spectrum tend to come from different sub-blocks, thus remain uncorrelated, as explained schematically in Fig.~\ref{fig:partition}. Only above the energy scale, $\delta_\mathrm{min} \equiv E_c$, two energy levels may come from the same spatial sub-block and become correlated. Hence, the \bte\ has a critical energy scale, $E_c$.

Upon unfolding, the energy becomes dimensionless keeping unit mean level spacing while the DOS becomes uniform~\cite{Guhr1998}. Then, the unfolded spectrum, $\cbr{\Ecal_j}$, has the critical energy scale
\begin{align}
	\Ecal_c \sim \frac{E_c}{\delta} = N^\gamma.
	\label{eq:E_c_unfold}
\end{align}
So, unfolded bulk energy levels with gap smaller than $\Ecal_c$ are necessarily uncorrelated as in the Poisson ensemble. The measures of short-range correlation (e.g.~ratio of level spacing~\cite{Atas2013}) exhibits a 2nd order phase transition at $\gamma = 0$, indicating that the short-range correlation is absent for $\gamma>0$ in the thermodynamic limit~\cite{Das2022}. However, two unfolded energy levels with gap larger than $\Ecal_c$ may be correlated as well as uncorrelated in case of the \bte, leading to an overall weak long-range correlation on average. 
Contrarily, in a typical many-body system and random matrix models like the Rosenzweig-Porter ensemble~\cite{Rosenzweig1960, Facoetti2016, Truong2016, Bogomolny2018, Soosten2019, Khaymovich2020, Biroli2021, Buijsman2022, DeTomasi2022, Sarkar2023}, the energy states hybridize with correlated energy levels forming a miniband~\cite{Altshuler2023} below the Thouless energy scale~\cite{Serbyn2017, Schiulaz2019, Corps2020, Sierant2020}. Consequently, the energy correlations are similar to that of the GOE below the Thouless energy scale. 
%above which the energy correlations cease to exist. 
Hence, the \bte\ exhibits an anomalous long-range correlation complimentary to that of a typical many-body system~\cite{Sierant2020a}.

%=================================
\subsection{Two-level form factor}\label{sec_b2}
%=================================
The long-range energy correlation can be extracted from the two-point energy correlation 
\begin{align}
    \mathcal{R}^{(2)} (\Ecal_1, \Ecal_2) \equiv \mean{\rho(\Ecal_1)\rho(\Ecal_2)}
\end{align}
where $\rho(\Ecal)$ is the DOS of unfolded spectrum. $\mathcal{R}^{(2)} (\Ecal_1, \Ecal_2)$ is the probability to find unfolded energy levels at $\Ecal_1$ and $\Ecal_2$. The two-point correlation has translational invariance, i.e.~$\mathcal{R}^{(2)}(\Ecal_1, \Ecal_2) = \mathcal{R}^{(2)}(\Egap)$ where $\Egap\equiv |\Ecal_1 - \Ecal_2|$. A related quantity is two-level cluster function, $Y_2(\Egap) \equiv 1 - \mathcal{R}^{(2)} (\Egap)$. For Poisson ensemble, $Y_2(\Egap) = 0$ at any energy gap.

%=================================
%	Two-level form factor
\begin{figure}[t]
	\centering
	\includegraphics[width=\columnwidth]{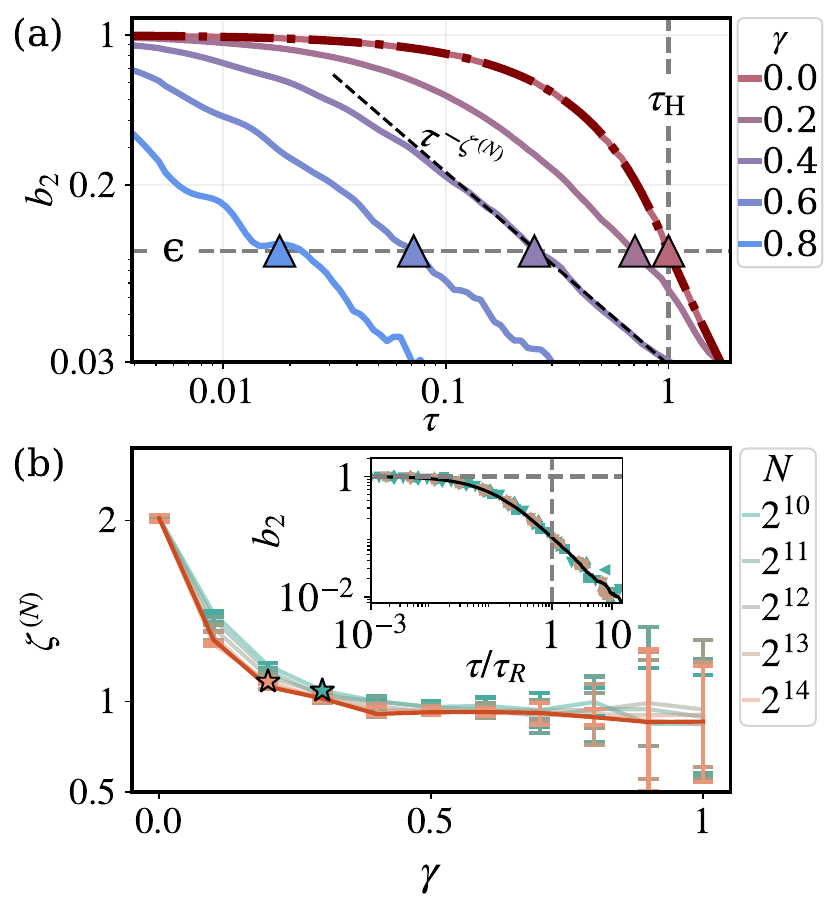}
	\caption{Two-level form factor. (a)~Evolution w.r.t.~dimensionless time, $\tau$ for $N = 1024$ and various values of $\gamma$. The analytical expression for GOE (Eq.~\eqref{eq:def_b2}) is shown via dashed lines. The vertical line denotes the dimensionless Heisenberg time, $\tauH \equiv 1$. The horizontal line denotes the threshold, $\epsilon$ (Eq.~\eqref{eq:tol_b2_def}) and the markers denote the relaxation time $\tauR$. The black dashed line shows the power law decay of $b_2(\tau>\tauR)$ for a fixed value of $\gamma$ according to Eq.~\eqref{eq:b2_beta}.
		%================
		(b)~Exponent of the asymptotic power-law decay of the two-level form factor, $b_2(\tau>\tauR) \propto \tau^{-\zeta^{(N)}}$, vs.~$\gamma$ for various system sizes. The thermodynamic ($N\to \infty$) extrapolation of the exponents is shown via bold curve. Asterisks indicate the analytically estimated values of $\gamma_{\ast}(N)$ (Eq.~\eqref{eq:gamma_star}) for the smallest ($N = 1024$) and the ($N = 16384$) system sizes. The corresponding collapse of $b_2(\tau/\tauR$) for $\gamma \gtrapprox \gamma_*(N)$  is shown in the inset.
	}
	\label{fig:b2}
\end{figure}
%=================================
%	b2 function
The Fourier transform of the two-level cluster function gives the two-level form factor in the dimensionless time domain, $\tau$
\begin{align}
	b_2(\tau) = \int_{-\infty}^{\infty} d\Egap\: Y_2(\Egap) \cos(2\pi \tau \Egap)
	\label{eq:def_b2}
\end{align}
where the complex integral reduces to a real one due to the symmetry $Y_2(\Egap) = Y_2(-\Egap)$~\cite{Buijsman2020}. The two-level form factor controls the long time dynamics of a generic observable~\cite{Schiulaz2019}. For Poisson ensemble, $b_2(\tau) = 0$ at any time while for GOE~\cite{Mehta2004book}
\begin{align}
	b_2^{\mathrm{GOE}}(\tau) &= \begin{cases}
		1 - 2\tau + \tau\log\del{1 + 2\tau}, & \tau\leq 1,\\
		\tau\log \left( \dfrac{2\tau + 1}{2\tau - 1} \right) - 1, &\tau > 1 .
	\end{cases}
	\label{eq:b2_GOE}
\end{align}

In the time evolution of a generic observable, the largest possible timescale is the Heisenberg time, defined as the inverse of the mean level spacing~\cite{Schiulaz2019, Kos2018}. For an unfolded spectrum, the mean level spacing is unity, hence, the dimensionless Heisenberg time, $\tauH = 1$ for any discrete energy spectrum. Consequently, we expect the two-level form factor to vanish at a sufficiently long time, i.e.~$b_2(\tau > \tauH) = 0$ to account for the absence of the energy correlation. However, the two-level form factor of GOE (Eq.~\eqref{eq:b2_GOE}) has a quadratic decay at long time, $b_2^{\mathrm{GOE}}(\tau\gg 1) \sim \tau^{-2}$. Then, we define a tolerance value as
\begin{align}
	\epsilon \equiv b_2^{\mathrm{GOE}}(\tauH) = \log(3) - 1 \approx 0.0986
	\label{eq:tol_b2_def}
\end{align}
such that for an arbitrary energy spectrum, whenever the two-level form factor falls below $\epsilon$, we expect the dynamics of a generic observable to equilibrate from a vanishing energy correlation. Then, the dimensionless time where the form factor reaches the tolerance value is the relaxation time $\tauR$, i.e.~$b_2(\tauR) = \epsilon$.

In Fig.~\ref{fig:b2}(a), we show the time evolution of the two-level form factor for various values of $\gamma$ and fixed system size, along with the tolerance value, $\epsilon$ from Eq.~\eqref{eq:tol_b2_def} shown via horizontal dashed lines. We identify the intersection of the two-level form factor with the tolerance limit from Eq.~\eqref{eq:tol_b2_def} as $\tauR$, which are shown via markers.

We observe that the two-level form factor exhibits a power-law decay at long time, $b_2(\tau > \tauR) \propto \tau^{-\zeta^{(N)}}$. We extract the finite-size power-law exponents for various values of $\gamma$, as shown in Fig.~\ref{fig:b2}(b) via solid lines with errorbars for various system sizes. From a linear extrapolation of the values of $\zeta^{(N)}$ w.r.t.~$N^{-1}$, we extract the intercepts i.e. the thermodynamic exponents, $\zeta\equiv \lim\limits_{N\to\infty} \zeta^{(N)}$ and show them in Fig.~\ref{fig:b2}(b)
as a function of $\gamma$ via bold line. It shows that the long-time power-law decay of the two-level form factor slows down as we go towards the localized regime. % Close to $\gamma = 1$, we find that $b_2(\tau > 1) \propto 1/\tau$ while $b_2(\tau > 1) \propto 1/\tau^2$ at $\gamma = 0$.
We also observe that away from $\gamma = 1$, $1-b_2(\tau)$ has a linear decay at times, small compared to the relaxation time. Since, the two-level form factor equilibrates at $\tau = \tauR$, we can approximate the behavior for $\gamma < 1$ as
\begin{align}
    b_2(\tau) \sim \begin{cases}
        1 - \dfrac{\tau}{c_\tau \tauR}, &\tau\ll \tauR\\
        \tau^{-\zeta}, &\tau > \tauR
    \end{cases}.
    \label{eq:b2_beta}
\end{align}
where $c_\tau \sim \mathcal{O}(1)$ is a function of $\gamma$ but independent of system size. In particular, $c_\tau = 1/2$ and $\zeta = 2$ for GOE ($\gamma = 0$). The linear behavior of $b_2(\tau)$ for $\tau\ll \tauR$ can be understood as follows: from the spatially local equivalence of the \bte\ to the Anderson model, we expect the typical energy correlations to be determined by the largest block, $\Delta_{l_\mathrm{max}}$. Corresponding energy spectrum is a superposition of $\mathcal{O}(N^\gamma)$ number of independent spectra following the same statistics as GOE, each coming from an independent sub-block. The resultant two-level cluster function is
\begin{align}
    Y_2^{\mathrm{mix}}(\Egap) = \sum_{k = 1}^{N^{\gamma}} f_k^2 Y_2^{(k)} (f_k \Egap)
\end{align}
where $Y_2^{(k)}(\Egap) = Y_2^{\mathrm{GOE}}(\Egap)$ and $f_k = N^{-\gamma}$ is the fraction of energy levels in a single sub-block. Thus, the two-level form factor becomes
\begin{equation}
\begin{aligned}
	b_2(\tau) &= \int_{-\infty}^{\infty} d\Egap\, N^{-\gamma} Y_2\left(\frac{\Egap}{N^{\gamma}}\right) \cos(2\pi \tau \Egap) \\
	          &= b_2^{\mathrm{GOE}}(N^{\gamma} \tau) \\
	          &\equiv b_2^{\mathrm{GOE}}\left(\frac{\tau}{2c_{\tau} \tau_{\mathrm{R}}}\right),
\end{aligned}
\label{eq:b2_mix}
\end{equation}
which corresponds to the asympototics given by Eq.~\eqref{eq:b2_beta}. From the local equivalence to 1D Anderson model the typical block of the \bte\ has a localization length $\xi \sim 8.75 N^{1-\gamma}$ (Eq.~\eqref{eq:xi_sub_block}). This localization length becomes comparable to the system size $N$ at 
\begin{align}
    \gamma_* (N) = \frac{\ln 8.75}{\ln N}.
    \label{eq:gamma_star}
\end{align}
For finite system sizes, the block Hamiltonian structure should be clearly visible for $\gamma > \gamma_*(N)$ and one should observe both $\zeta$ and $c_\tau$ to be independent of $\gamma$ and $N$. We show that the two-level form factor collapses as a function of $\tau/\tau_R$ and exhibits a universal behavior for $\gamma > \gamma_{\ast}(N)$ , as shown in the inset of Fig.~\ref{fig:b2}(b). Note that, Eq.~\eqref{eq:gamma_star} estimates the finite size effect while the block picture is valid for any $\gamma>0$ in the thermodynamic limit.

%=================================
\subsection{Number variance}\label{sec:num_var}
%=================================
The spatially local equivalence of the \bte\ to the 1D Anderson model (Fig.~\ref{fig:partition}) shows a multi-scale local spectra manifesting into an atypical long-range correlation in the global energy spectrum. Corresponding critical energy, $\Ecal_c$ (Eq.~\eqref{eq:E_c_unfold}) is reflected in the heterogeneous behavior of the measures of long-range energy correlations, e.g.~number variance~\cite{Baecker1995},
\begin{align}
	\begin{split}
		\Sigma^2(\Egap) &= \mean{ [\mathcal{N}(\Ecal_0, \Egap) - \Egap]^2}_{\Ecal_0}\\
		\mathcal{N}(\Ecal_0, \Egap) &= I\del{\Ecal_0 + \frac{\Egap}{2}} - I\del{\Ecal_0 - \frac{\Egap}{2}}
	\end{split}
	\label{eq:nvar_def}
\end{align}
where $I(\Ecal)$ is the cumulative distribution of the unfolded energy levels, $\mathcal{N}(\Ecal_0, \Egap)$ is the number of energy levels in the window of span $\Egap$ and centered at $\Ecal_0$. The number variance is related to the two-point energy correlation as~\cite{Mehta2004book}
\begin{align}
	\Sigma^2(\Egap) = \Egap - 2 \int_{0}^{\Egap} dx(\Egap - x) Y_2(x) .
	\label{eq:nvar_Y2_def}
\end{align}
For GOE, the energy spectrum is rigid so that any two energy levels are correlated with a logarithmic number variance~\cite{Mehta2004book}

\begin{align}
    \Sigma_{\mathrm{GOE}}^2(\Egap) \approx
    \frac{2}{\pi^2}\left( \ln(2\pi\Egap) + \gEM + 1 -\frac{\pi^2}{8}\right)\label{eq:nvar_GOE}
\end{align}
where $\gEM$ is the Euler-Mascheroni constant.
In the Poisson ensemble, the number of unfolded energy levels are free to fluctuate around their mean position, $\mean{\mathcal{N}(\Ecal_0, \Egap)}_{\Ecal_0} = \Egap$ and the number variance exhibits a linear behavior, $\Sigma^2(\Egap) = \Egap$.

%=================================
\iffalse
\begin{align}
	\label{eq_num_var_GOE}
	\begin{split}
		\Sigma^2_\mathrm{GOE}(\Ecal) &= \frac{2}{\pi^2}\del{ 1 + \gEM + \log(2\pi\Ecal) + \pi^2 \Ecal - \cos (2\pi\Ecal) - \mathrm{Ci}(2\pi\Ecal) - 2\pi\Ecal \mathrm{Si}(2\pi\Ecal) - \frac{\pi}{2}\mathrm{Si}(\pi\Ecal) + \frac{\mathrm{Si}^2(\pi\Ecal)}{2} }
	\end{split}\nonumber\\
	&= \frac{2}{\pi^2} \log\del{ 2\pi \Ecal + \gEM + 1 - \frac{\pi^2}{8} } + \mathcal{O}\del{\frac{1}{\Ecal}}, \qquad \gEM = 0.577216\dots
\end{align}
where $\gEM$ is the Euler-Mascheroni constant, $\mathrm{Ci}(x) = -\int_{x}^{\infty}dt \frac{\cos t}{t}$ and $\mathrm{Si}(x) = \int_{0}^{x}dt \frac{\sin t}{t}$.
\fi
%============================

\begin{figure}[t]
	\centering
	\includegraphics[width=\columnwidth]{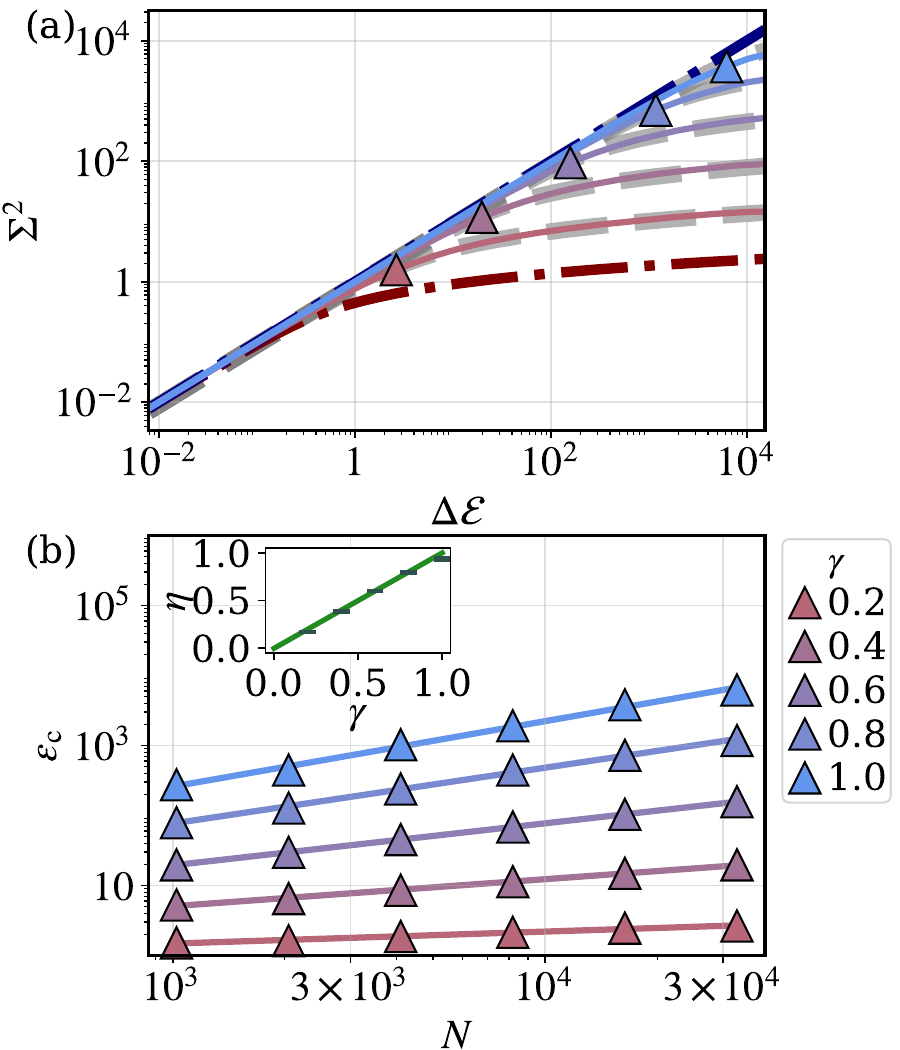}
	\caption{Long-range energy correlation. (a)~Number variance for $N = 32768$ and different values of $\gamma$. The red (blue) dashed curve describes the GOE (Poisson ensemble). The markers denote the critical dimensionless energy, $\Ecal_c$ (Eq.~\eqref{eq:E_c_unfold}). The gray dashed lines correspond to the analytical prediction for $\Egap > \Ecal_c$ of (Eq.~\eqref{eq:nvar_beta}).
		%=========================
		(b)~Critical energy vs.~system size, $N$ for various values of $\gamma$ where the solid lines denote linear fit in the log-log scale. Inset shows the system size scaling exponent of the critical energy, i.e.~$\Ecal_c \propto N^\eta$ where the solid line denotes $\eta = \gamma$. Data are averaged over 1024 disorder realizations. Error-bars in the insets denote 95\% confidence interval. 
	}
	\label{fig:num_var}
\end{figure}

In Fig.~\ref{fig:num_var}(a), we show the ensemble averaged number variance of the \bte\ for $N = 32768$ and various values of $\gamma$ along with the analytical expressions of the GOE and Poisson ensemble. For sufficiently small energy interval $\Egap$, $\Sigma^2(\Egap)$ has a linear behavior as in the Poisson ensemble. Upon increasing the energy interval, the number variance deviates from the Poisson behavior and indicates the presence of long-range correlation. In Fig.~\ref{fig:num_var}(a), we show that the inverse of the relaxation time $\tauR$ obtained from the two-level form factor (Fig.~\ref{fig:b2}(a)) determines the critical energy, $\Ecal_c$, above which the long-range correlation exists. Such a critical energy can be clearly identified from the heterogeneous behavior of the power spectrum of noise~\cite{Das2022, Das2023}, also useful to identify the Thouless energy scale~\cite{Berkovits2020, Corps2020}.

The system size dependence of the critical energy is shown in Fig.~\ref{fig:num_var}(b) in log-log scale, which indicates a power-law dependence, i.e.~$\Ecal_c \propto N^\eta$ where the proportionality constant is $\mathcal{O}(1)$ and independent of $\gamma$. The inset of Fig.~\ref{fig:num_var}(b) shows that the power-law exponent, $\eta \approx \gamma$ and validates Eq.~\eqref{eq:E_c_unfold}, i.e.~$\Ecal_c\propto N^\gamma$. Thus, dimensionless relaxation time has the following behavior
\begin{align}
	\tauR \approx \Ecal_c^{-1} \propto N^{-\gamma}.
	\label{eq:tauR_beta}
\end{align}
Above relation is a consequence of the Fourier duality between energy and time, where the absence of energy correlation for $\Egap < \Ecal_c$ translates into the relaxation of dynamics for $\tau > \tauR$. Eq.~\eqref{eq:tauR_beta} implies that in the NEE phase of the \bte, the relaxation time is much smaller than the Heisenberg time and decreases with the system size.

%=================================
%	Number variance
%=================================

From the schematic in Fig.~\ref{fig:partition}, we see that similar to the two-level cluster function, the typical behavior of the number variance is also governed by the largest block $\Delta_{l_\mathrm{max}}$. For any energy interval of width $\Egap < \Ecal_c$, the number variance grows linearly due to the contribution of $\mathcal{O}(1)$ levels from each of the $\mathcal{O}(N^\gamma)$ independent sub-blocks. However, at larger energy intervals ($\Egap > \Ecal_c$), each of the sub-block contributes $\mathcal{O}(\Egap/\Ecal_c)\gg 1$ number of correlated energy levels as in the GOE. Such a superposition of uncorrelated spectra has the number variance
\begin{align}
    \Sigma^{2}_{\mathrm{mix}}(\Egap) = \sum_{k = 1}^{N^{\gamma}} \Sigma^{2}_{(k)} (f_k \Egap)
\end{align}
where $\Sigma^{2}_{(k)}(\Egap) = \Sigma^{2}_{\mathrm{GOE}}(\Egap)$ and $f_k = N^{-\gamma}$. Combining the short- and long-range behaviors discussed above, we get the number variance of the \bte\ as~\cite{Ossipov_Sigma2_GOE_PLRBM}
\begin{align}
	\Sigma^2(\Egap) \sim \begin{cases}
		\Egap, & \Egap \lesssim \Ecal_c\\
		\dfrac{\Ecal_c}{\pi^2}\left( \ln \dfrac{2\pi\Egap}{\Ecal_c} \right), & \Egap \gtrsim \Ecal_c %+\gamma_{EM}+1
	\end{cases}
    \label{eq:nvar_beta}
\end{align}
where $\Ecal_c \sim N^\gamma$ is the unfolded critical energy (Eq.~\eqref{eq:E_c_unfold}). 
%where $\gamma_{EM}\approx 0.57$ is the Euler-Mascheroni constant and all the prefactors and offsets taken from the GOE result, see, e.g., Eq.(6) in~\cite{Ossipov_Sigma2_GOE_PLRBM}.
In Fig.~\ref{fig:num_var}(a), we plot the numerically obtained number variance averaged over ensemble and spectrum along with the analytical prediction from Eq.~\eqref{eq:nvar_beta} via dashed gray curves for different values of $\gamma$ and find an excellent agreement.

\iffalse
We find that in the NEE phase of the \bte, the number variance shows a power-law behavior above the critical energy scale. For a fixed system size, $N$, we extract such power-law exponents as $\Sigma^2(\Egap > \Ecal_c) \propto \Egap^{\eta^{(N)}}$. From a linear fit of such finite size exponents w.r.t.~$N^{-1}$, we extract the thermodynamic behavior, $\lim\limits_{N\to \infty}\eta^{(N)} = \eta$. In the inset of Fig.~\ref{fig:num_var}(a), we show the extrapolated power-law exponent, $\eta$ as a function of $\gamma$, which indicates that $\eta \approx \gamma$. Thus, long-range energy correlation in the NEE phase behaves as
\begin{align}
	\Sigma^2(\Egap) \propto \begin{cases}
		\Egap, & \Egap < \Ecal_c\\
		\Egap^\gamma, & \Egap > \Ecal_c
	\end{cases}.
	%\label{eq:nvar_beta}
\end{align}
\fi

So, a weak long-range energy correlation persists despite the absence of any short-range correlation in the NEE regime of the \bte. In contrast, the number variance of a typical many-body system away from the localized phase behaves as in the Rosenzweig-Porter ensemble~\cite{Efetov1983}
\begin{align}
	\Sigma^2(\Egap) \propto \begin{cases}
		\log \Egap, & \Egap < \Ecal_\mathrm{Th}\\
		\Egap, & \Egap > \Ecal_\mathrm{Th}
	\end{cases}
	\label{eq:nvar_typ}
\end{align}
such that the energy levels are correlated below the Thouless energy scale, $\Ecal_\mathrm{Th}$ and uncorrelated otherwise. Comparing Eqs.~\eqref{eq:nvar_beta} and \eqref{eq:nvar_typ}, we observe that the critical energy $\Ecal_c$ of the \bte\ behaves complimentary to $\Ecal_\mathrm{Th}$ and present an anomalous long-range energy correlation. Similar anomalous critical energy is also expected in constrained matrix models with correlated entries~\cite{Mondal2017}.

In the next section, we look at the spectral form factor, whose non-equilibrium features are controlled by the long-range energy correlations discussed so far.

%=================================
\section{Spectral form factor}\label{sec:SFF}
%=================================

%=================================
% \subsection{SFF}
%=================================
%	SFF
\begin{figure*}[t]
	\centering
	\includegraphics[width=1.5\columnwidth]{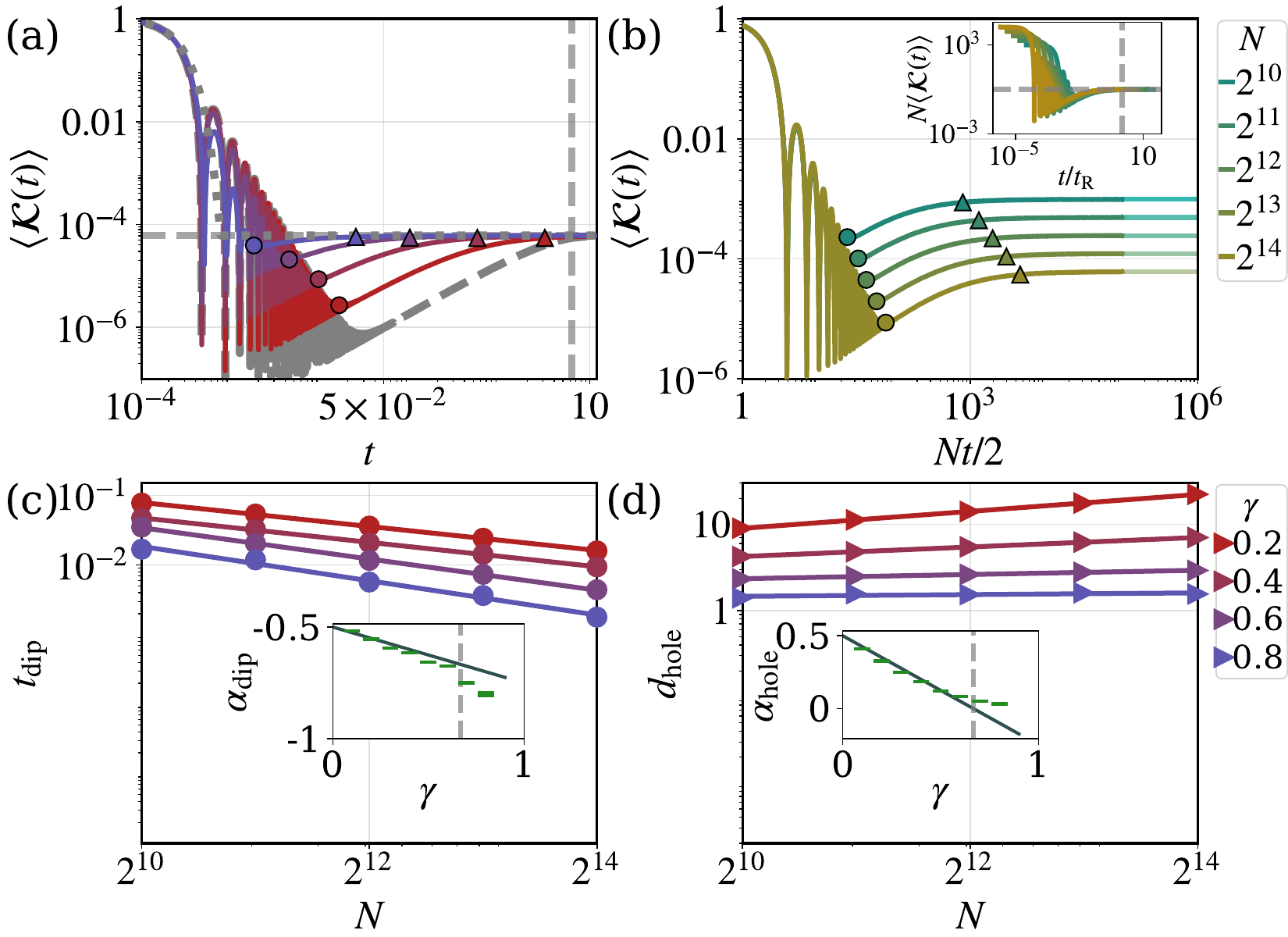}
	\caption{Spectral form factor. (a)~Time evolution of the ensemble averaged SFF for $N = 16384$ and various values of $\gamma$. The gray dashed (dotted) curve denotes the SFF of GOE (Poisson ensemble) given in Eq.~\eqref{eq:SFF_GOE} (Eq.~\eqref{eq:SFF_Psn}). The horizontal (vertical) dashed line denotes the asymptotic equilibrium value, $N^{-1}$ (Heisenberg time $\tH^{\mathrm{GOE}} = 2\pi$). The triangular (circular) markers denote the relaxation (dip) time, $\tR$ ($\tdip$).
		%=====================
		(b)~SFF for $\gamma = 0.4$ and various system sizes with time axis scaled as $Nt/2$. Markers denote the characteristic timescales as in panel (a).
		The inset shows the collapse of the ramp of $N$\sff{t} w.r.t. $t/\tR$.
        %=====================
		(c)~Dip time, $\tdip\propto N^{\alpha_\mathrm{dip}}$ and (d)~relative depth of the correlation hole, $d_\mathrm{hole} \propto N^{\alpha_\mathrm{hole}}$ vs.~system size in log-log scale for various values of $\gamma$ along with linear fit. The insets of panel (c) and (d) show $\alpha_\mathrm{dip} \approx -\frac{1}{2} - \frac{\gamma}{4}$ and $\alpha_\mathrm{hole} \approx \frac{1}{2} - \frac{3\gamma}{4}$, respectively, while grey vertical dashed line denotes $\gamma_{\mathrm{T}} =2/3$ (Eq.~\eqref{eq:gamma_th}).}
	\label{fig:SFF}
\end{figure*}
%=================================
The SFF can be defined in terms of the energy levels $\left\{ E_n \right\}$ as
\begin{align}
	\begin{split}
		\sff{t} &= \frac{\abs{\sum\limits_{n = 1}^{N} e^{-iE_n t}}^2}{N^2} = \frac{1}{N^2} \sum_{n, m = 1}^{N} e^{-i(E_m - E_n)t}.
	\end{split}
	\label{eq:SFF_def}
\end{align}
Then, the ensemble averaged SFF can be expressed as~\cite{Mehta2004book, Schiulaz2019, Das2025b}
\begin{align}
	\begin{split}
		\mean{\sff{t}} &\approx \abs{\int dE\; e^{-iEt}\mean{\rho(E)}}^2 + \frac{1}{N} \left(1 - b_2\del{\frac{t}{\tH}} \right )
	\end{split}
	\label{eq:SFF_parts}
\end{align}
where the first term is the disconnected SFF, $\mathcal{K}_{\mathrm{d}}(t)$ and the last term is the connected SFF, $\mathcal{K}_c(t)$. In Eq.~\eqref{eq:SFF_parts}, $\mean{\rho(E)}$ is the limiting DOS, $b_2(t)$ is the two-level form factor. The disconnected part of the SFF determines the initial decay, {\itshape slope} while the equilibrium value, {\itshape plateau}, is $N^{-1}$. The SFF asymptotically approaches the plateau at the relaxation time, $\tR$.

The connected part of the SFF reflects the energy correlations and forces the SFF to be below the plateau at an intermediate time window, known as the {\itshape ramp} or {\itshape correlation hole}. The interplay between the slope and ramp part of the SFF leads to the minimum, which is called the {\itshape dip}, observed at 

\begin{align}
    \textrm{dip time} \equiv \tdip.
\end{align}
Hence, the SFF in a generic system exhibits a slope-dip-ramp-plateau structure~\cite{Das2025}. Such features are most prominent in case of a chaotic system, e.g.~the ensemble averaged SFF of GOE is~\cite{Das2025b, Kumar2025Arxiv}
\begin{align}
	\mean{\mathcal{K}^{\mathrm{GOE}}(t)} = \frac{16\fbslj{1}{\frac{N t}{2}}^2}{N^2t^2} - \frac{1}{N} b_2^{\mathrm{GOE}}\del{ \frac{t}{\tH^{\mathrm{GOE}}}} + \frac{1}{N},
	\label{eq:SFF_GOE}
\end{align}
where $b_2^{\mathrm{GOE}}(t)$ is given in Eq.~\eqref{eq:b2_GOE} and $\tH^{\mathrm{GOE}} = 2\pi$. The SFF of GOE exhibits a prominent correlation hole, as shown in Fig.~\ref{fig:SFF}(a). Contrarily, in a uncorrelated spectrum, the correlation hole is absent, e.g.~the average SFF of the Poisson ensemble is~\cite{TorresHerrera2014, TorresHerrera2014a, Tavora2016}
\begin{align}
	\begin{split}
		\mean{\mathcal{K}^{\mathrm{P}}(t)} &\approx \exp\del{- \frac{N^2t^2}{16}} + \frac{1}{N} ,% &= \Theta\del{\tR^{\textrm{P}} - t} e^{-\sigma_E^2 t^2} + \Theta\del{t - \tR^{\textrm{P}}} \frac{1}{N} ,
	\end{split}
	\label{eq:SFF_Psn}
\end{align}
%where $\Theta(x)$ is the Heaviside step function and 
where following the initial Gaussian decay, the SFF equilibrates at $\tR^{\textrm{P}}= 4\sqrt{\ln N}/N$ without forming any correlation hole. Therefore, a dimensionless dynamical measure of the energy correlation is the relative depth of the correlation hole
\begin{align}
	d_\mathrm{hole} = \frac{\sff{\tR}}{\sff{\tdip}} = \frac{1}{N \sff{\tdip} }
	\label{eq:corr_depth}
\end{align}
as well as the relative width of the correlation hole

\begin{align}
	w_\mathrm{hole} = \frac{\tR}{\tdip}
	\label{eq:corr_width}
\end{align}
where $\tdip$ denotes the intersection of the slope and ramp part of the SFF. For localized systems, the correlation hole is absent, so both the relative depth and width of correlation hole is unity. In contrast, for GOE, the dip time is $\tdip^\mathrm{GOE} \approx 4\frac{3^{\frac{1}{4}} }{\sqrt{2N}}$ where the SFF has a value $\mathcal{K}^\mathrm{GOE}(\tdip) \sim N^{-\frac{3}{2}}$~\cite{Das2025b}. Then, using Eqs.~\eqref{eq:corr_depth} and \eqref{eq:corr_width}, we get the relative depth and width of the correlation hole for GOE as
\begin{align}
    \begin{split}
        d_\mathrm{hole}^\mathrm{GOE} &\sim N^{\frac{1}{2}},\quad w_\mathrm{hole}^\mathrm{GOE} \sim N^{\frac{1}{2}}.
    \end{split}
    \label{eq:hole_GOE}
\end{align}
Thus, for a correlated spectrum, both the relative depth and width of the correlation hole increases with system size. Both the quantities approach unity as we approach the delocalization-localization transition, beyond which the energy correlation is absent at all possible length scales.

In Fig.~\ref{fig:SFF}(a), we show the ensemble averaged SFF of the \bte\ for a fixed system size and various values of $\gamma$ along with the analytical expressions of the GOE (Eq.~\eqref{eq:SFF_GOE}) and Poisson ensemble (Eq.~\eqref{eq:SFF_Psn}). The Heisenberg time for GOE, $\tH^\mathrm{GOE} = 2\pi$ is shown via vertical dashed line in Fig.~\ref{fig:SFF}(a), while the relaxation times for different $\gamma$ values are shown via marker. Our analyses of the dimensionless relaxation time ($\tauR\propto N^{-\gamma}$; Eq.~\eqref{eq:tauR_beta}, Fig.~\ref{fig:b2}(a)) obtained from the two-level form factor in the NEE phase of the \bte, the relaxation time of the SFF scales as
\begin{align}
    \tR \propto N^{-\gamma}
    \label{eq:tR_scaling}
\end{align}
since $\tH \sim \mathcal{O}(1)$. Hence, the relaxation time decreases as we approach the localized regime. In Fig.~\ref{fig:SFF}(b), we present the SFF for $\gamma = 0.4$ across different system sizes, with the time axis scaled as $Nt/2$ to ensure a collapse of the initial oscillations.

%================================= 
%Disconnected part of 2-point SFF
\begin{figure}[t]
    \centering
    \includegraphics[width=\columnwidth]{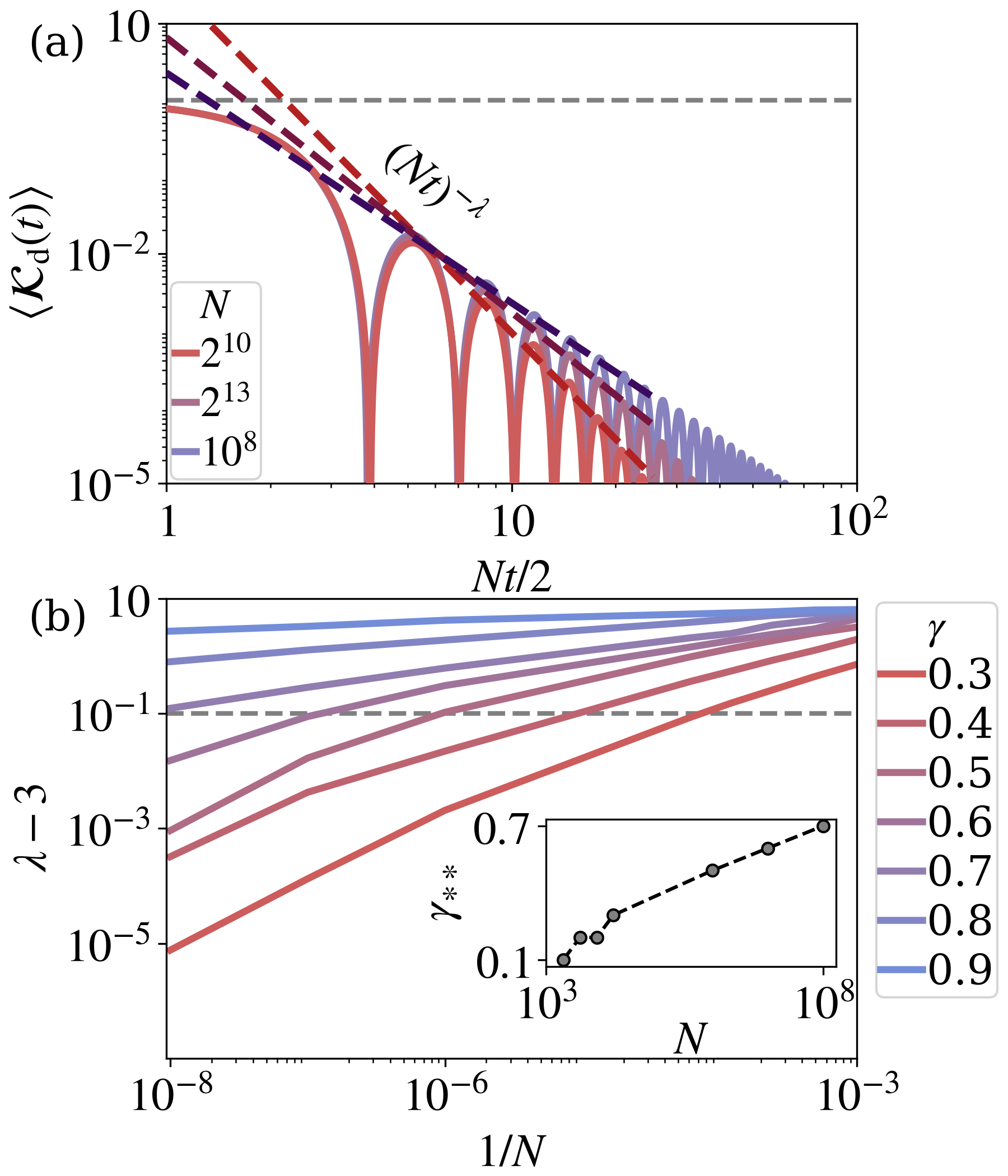}
    \caption{Disconnected part of SFF. (a)~Time evolution of the ensemble averaged disconnected SFF of (Eq.~\eqref{eq:SFF_parts}) for $\gamma = 0.5$ and different values of $N$. The time axis is scaled by a factor of $N/2$. The dashed lines correspond to the power-law decays ($\sim (Nt)^{-\lambda}$).
    %================
		(b)~Power-law exponent \(\lambda\) subtracted by 3, plotted against \(1/N\) for various values of \(\gamma\). The horizontal dashed line corresponds to a threshold value of 0.1. In inset circular markers denote $\gamma_{**}(N)$ estimated using Eq.~\eqref{eq:gamma_star_star}.}
    \label{fig:SFF_disconnected}
\end{figure}
%=================================

In Fig.~\ref{fig:SFF_disconnected}(a) we plot the ensemble averaged disconnected SFF as given in Eq.~\eqref{eq:SFF_parts} for a fixed value of $\gamma$ and different system sizes. We show the decay of the envelope via dashed lines. The initial decay, i.e.~slope of the SFF ($\sim (Nt)^{-\lambda})$ is dictated by the scaling at the edge of the DOS~\cite{Urbanowski2009, Khalfin1958, Tavora2016, TorresHerrera2014a, ZarateHerrada2025arxiv}. In case of the GOE, the slope of the SFF exhibits a power-law decaying envelope, $\sim (Nt)^{-3}$~\cite{Schiulaz2019}. In case of the \bte, the DOS exhibits a second order phase transition at $\gamma = 1$ from semicircle law to Gaussian distribution~\cite{Das2023}. Hence, in the thermodynamic limit, the DOS of the \bte\ follows Wigner's semicircle law in the entire nonergodic regime, yielding $\sim (Nt)^{-3}$ decay of the slope independent of $\gamma$, as shown in Fig.~\ref{fig:SFF_disconnected}(b) where we plot $\lambda - 3$ as a function of $1/N$ in log-log scale to see the convergence of $\lambda$ towards $3$. At the localization transition point, $\gamma = 1$, the DOS is system size invariant such that the initial decay of the SFF is independent of $N$. Moreover, the two-level form factor, $b_2(\tau)$ collapses for different system sizes w.r.t.~$\tau/\tauR$ such that ramp of the SFF collapse w.r.t.~$t/\tR$ as shown in the inset of Fig.~\ref{fig:SFF}(b). Such scale-invariant dynamics is previously observed at the critical point of quasi-periodic and disordered single particle models~\cite{Hopjan2023} as well as interacting many-body systems~\cite{Hopjan2023a}.

The two-level form factor given in Eq.~\eqref{eq:b2_beta} implies that the ramp of the SFF has a linear growth at small time, $t \ll \tR$. Since the slope ($\sim (Nt)^{-3}$) and ramp intersects at the dip time, we can estimate the same as
\begin{align}
    (N\tdip)^{-3} \sim \frac{1-b_2(\tdip)}{N} \Rightarrow \tdip \sim N^{-\frac{1}{2} - \frac{\gamma}{4}}
    \label{eq:tdip_scaling}
\end{align}
where the SFF has a value 
\begin{align}
    \sff{\tdip} \sim (N\tdip)^{-3} \sim N^{\frac{3\gamma - 6}{4}}.
    \label{eq:SFF_tdip_scaling}
\end{align}
Then, the relative depth of the correlation hole scales as,
\begin{align}
    d_\mathrm{hole}\sim \frac{1}{N\sff{\tdip}} = N^{\frac{1}{2} - \frac{3\gamma}{4}}
    \label{eq:d_hole_pred}
\end{align}
while the relative width scales as
\begin{align}
    w_\mathrm{hole}\sim \dfrac{\tR}{\tdip} = N^{\frac{1}{2} - \frac{3\gamma}{4}}.
    \label{eq:w_hole_pred}
\end{align}
Both of these scalings are consistent with the GOE limit given in Eq.~\eqref{eq:hole_GOE}. 
In Fig.~\ref{fig:SFF}(c) and (d), we respectively show the dip time, $\tdip$ and the relative depth of the correlation hole as a function of system size for various values of $\gamma$ with the scaling exponents given in the insets as a function of $\gamma$. We find that close to $\gamma = 0$, Eqs.~\eqref{eq:tdip_scaling}, \eqref{eq:d_hole_pred} and \eqref{eq:w_hole_pred} captures the system size scaling of the dip time, relative depth and width of the correlation hole. However, the deviations from our analytical predictions close to $\gamma = 1$ indicate a strong finite size effect.

It can be noted from Fig.~\ref{fig:SFF_disconnected}(b) that depending on $N$, there exists a point $\gamma_{**}(N)$ beyond which the decay of the envelope is not $(Nt)^{-3}$ due to the finite size effect. We estimate $\gamma_{**}(N)$ as 
\begin{align}
    \lambda(N)-3 > \varepsilon,~~\forall~\gamma> \gamma_{**}(N)
    \label{eq:gamma_star_star}
\end{align}
where $\varepsilon$ is the threshold value which we set as 0.1.
The estimated values of $\gamma_{**}(N)$ are shown via circular markers in the inset of Fig.~\ref{fig:SFF_disconnected}(b). However, in the thermodynamic limit $\gamma_{**}(N) \rightarrow 1$ in the entire NEE phase.

\begin{figure}[t]
	\centering
	\includegraphics[width=\columnwidth]{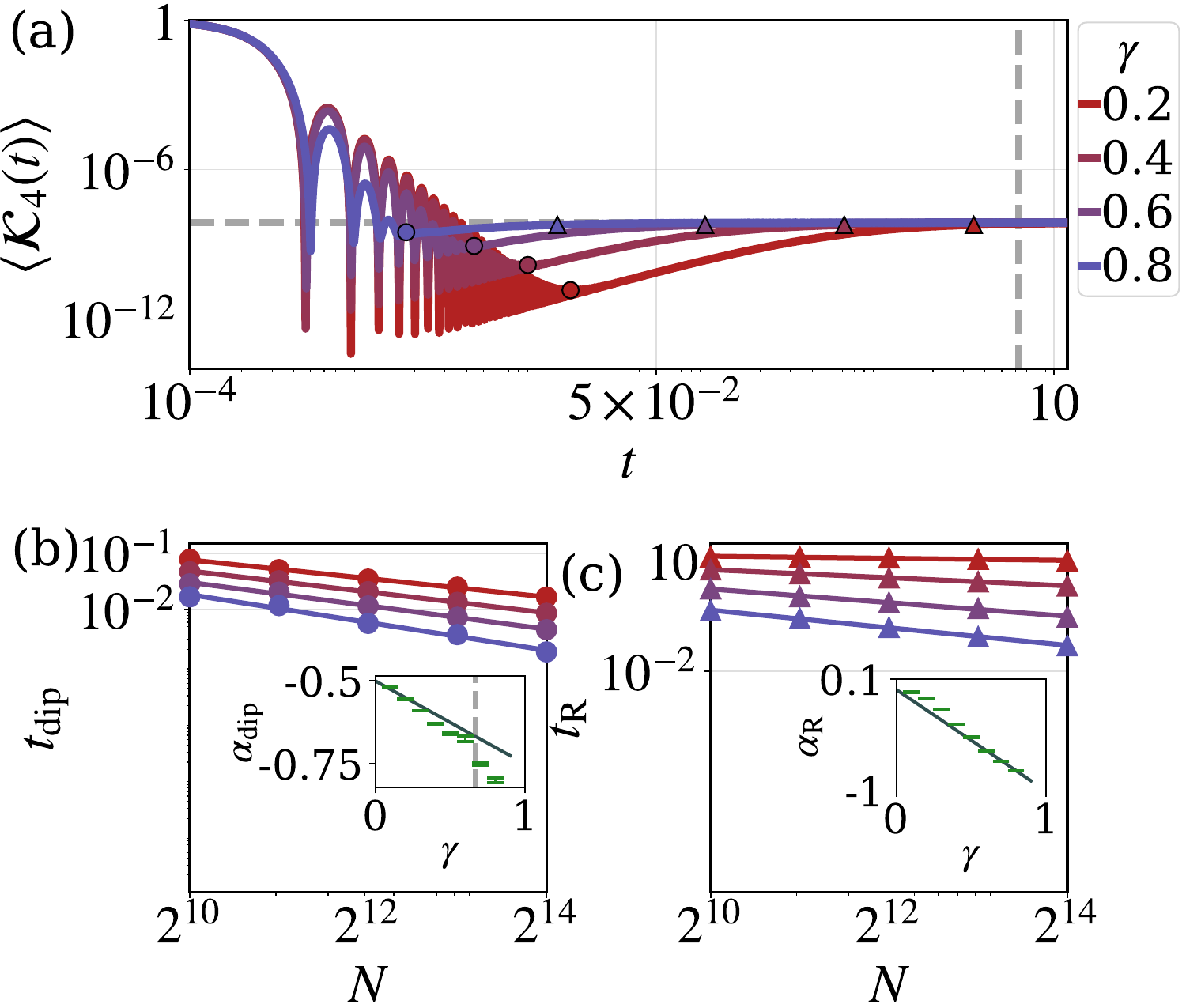}
	\caption{4-point SFF (Eq.~\eqref{eq:SFF_4_def}). (a)~ensemble averaged 4-point SFF for $N = 16384$ and various values of $\gamma$. The horizontal dashed line denotes the asymptotic equilibrium value, $2/N^2$ and the vertical dashed line denotes $\tH^{\mathrm{GOE}}$. The triangular (circular) markers denote the relaxation (dip) time.
		%=====================
		(b)~dip time, $\tdip \propto N^{\alpha_\mathrm{dip}}$ and (c)~relaxation time, $\tR\propto N^{\alpha_\mathrm{R}}$ vs.~system size in log-log scale for various values of $\gamma$ along with linear fit while the inset shows the scaling exponents as a function of $\gamma$. The grey vertical dashed line in the inset of panel (b) denotes $\gamma_{\mathrm{T}} =2/3$ (Eq.~\eqref{eq:gamma_th}). 
	}
	\label{fig:SFF_4}
\end{figure}

The relaxation time $\tR$ is always expected to be larger than the dip time $\tdip$ and $\sff{\tdip}$ must be less than the asymptotic value of the SFF $1/N$, therefore Eqs.~\eqref{eq:tR_scaling}, \eqref{eq:tdip_scaling}, and~\eqref{eq:SFF_tdip_scaling}  are valid only for $\gamma<2/3$. From Fig.~\ref{fig:SFF}(a) it is observed that because of the interplay of the connected and disconnected SFF parts  the ramp is hidden behind the initial slope part after this certain value of $\gamma$, 
\begin{align}
    \gamma_{\mathrm{T}} = \frac{2}{3}.
    \label{eq:gamma_th}
\end{align}
  This threshold is also marked in the insets of Fig.~\ref{fig:SFF}(c), \ref{fig:SFF}(d) and \ref{fig:SFF_4}(b).
However, it is important to emphasize that $\gamma_{\mathrm{T}}$ is spurious and does not correspond to any genuine phase transition and the connected part itself still shows the ramp behavior, see Fig.~\ref{fig:b2}~\footnote{This is also visible in the total SFF at the minima of the oscillations of the disconnected SFF part, see, e.g., Fig.~\ref{fig:SFF}.}.
We find that both $d_\mathrm{hole}$ and $w_\mathrm{hole}$ shrink in the NEE phase of the \bte\ as we increase $\gamma$. So, monitoring the relative depth and width of the correlation hole of the SFF, we can detect the localization transition point of the \bte, i.e.~$\gamma = 1$.

%%=================================
%\subsection{4-point SFF}
%%=================================
%=================================
%	4-point SFF
%=================================
Equation~\eqref{eq:SFF_def} implies that the SFF is a two-point correlation function. In a similar manner, one can define the ensemble averaged 4-point SFF as~\cite{Liu2018, Cotler2017}
\begin{align}
    \begin{split}
	\mean{\mathcal{K}_4(t)} &=\frac{\abs{\sum\limits_{n = 1}^{N} e^{-iE_n t}}^4}{N^4}.
	\label{eq:SFF_4_def}
    \end{split}
\end{align}

The 4-point SFF is related to the out-of-time order correlator of a generic observable in case of a chaotic system. Hence, the short-time decay of the 4-point SFF is important for understanding early-time chaos, for example, the Lyapunov exponent obtained from the growth rate of the out-of-time order correlator. 

In Fig.~\ref{fig:SFF_4}(a), we show the time evolution of the 4-point SFF for various values of $\gamma$, where the equilibrium value is $2/N^2$~\cite{Liu2018}. We observe a slope-dip-ramp-plateau structure of the 4-point SFF similar to its two-point counterpart in the NEE phase of \bte. We also observe that the characteristic timescales of the 4-point SFF behave similar to their two-point counterparts
\begin{align}
	\tdip \propto N^{-\frac{1}{2} - \frac{\gamma}{4}},\quad \tR \propto N^{-\gamma}
\end{align}

as shown in Fig.~\ref{fig:SFF_4}(b) and (c), respectively.

%=================================
\section{Discussion}\label{sec:discuss}
%=================================
In this work, we investigate the dynamical properties of the \bte, focusing on how anomalous energy correlations manifest in the number variance and the spectral form factor (SFF). In the nonergodic phase of the \bte, a spatially local equivalence to the 1D Anderson model reveals a multi-scale spectrum and a critical dimensionless energy $\Ecal_c \sim N^\gamma$, where $N$ is the system size and the Dyson index $\beta = N^{-\gamma}$. The number variance, $\Sigma^2(\Egap)$, deviates from the Poisson behavior at $\Ecal_c$ and we verify its system size scaling. The critical energy $\Ecal_c$ behaves complimentary to the Thouless energy scale observed in typical many-body systems: weak long-range correlations 
($\Sigma^2(\Egap) \sim \Ecal_c \ln (2\pi\Egap/\Ecal_c)$) exist above $\Ecal_c$ but energy correlations are absent below $\Ecal_c$ (Fig.~\ref{fig:Schematic_SFF}). We numerically obtain the two-level form factor as a function of the dimensionless time $\tau$ and show that the dimensionless relaxation time $\tauR$ is the inverse of $\Ecal_c$.

\iffalse
\AKD{In a typical many-body system, dynamical quantities like SFF equilibrate at the Heisenberg time, $\tH$. However, in the nonergodic phase of the \bte, the time evolution of a generic observable equilibrates at the relaxation timescale $\tR = \tauR\times \tH$, which is parametrically smaller than $\tH$. We call the temporal window between $\tR$ and $\tH$ as the {\corrvoid}, which reflects the absence of short-range correlation. Below $\tR$, the SFF exhibits a correlation hole reflecting the presence of long-range correlation. We show that by monitoring the gradual depletion of the correlation hole, we can identify the localization transition at $\gamma = 1$. 
%relative depth and width of the shrink as we approach the localized regime. 
We also look at the 4-point SFF, whose non-equilibrium features and scaling of the characteristic timescales are similar to its two-point counterpart.

Our findings highlight the role of energy correlations in the equilibration of a disordered system in the nonergodic phase. The anomalous critical energy found in the \bte\ is also expected to be present in other random matrix models like the banded matrices~\cite{Das2025b}, Burin-Maksimov model~\cite{Tang2022}, constrained ensembles with correlated entries~\cite{Mondal2017} and physical systems like kicked rotor~\cite{Pandey2017, Kumar2020}, complex atoms~\cite{Chirikov1985}, isolated thick wire~\cite{Iida1990, Fyodorov1993}.}
\fi

In a typical non-integrable interacting many-body system, the energy correlations follow Eq.~\eqref{eq:nvar_typ} away from the many-body localization. As a result, the ensemble averaged SFF exhibits a universal behavior dictated by GOE above the Thouless time~\cite{Kos2018, Fritzsch2024}. Beyond the Heisenberg time, the discreteness of the energy spectrum ensures the equilibration of the ensemble averaged SFF such that the relaxation time coincides with $\tH$. This is schematically explained in Fig.~\ref{fig:Schematic_SFF}(a), where we show the connected SFF (Eq.~\eqref{eq:SFF_parts}). As the many-body localized phase is approached, the Thouless energy gets smaller such that the Thouless time increases and eventually merges with the Heisenberg time such that the correlation hole vanishes. Such a behavior is observed in various spin-chains~\cite{Schiulaz2019, TorresHerrera2014}, 3D Anderson model~\cite{Sierant2020}, and random matrix models~\cite{Hopjan2023a}.

Contrarily, in the NEE phase of the \bte, the time evolution of a generic observable equilibrates at the relaxation timescale $\tR$, which is parametrically smaller than $\tH$, as schematically shown in Fig.~\ref{fig:Schematic_SFF}(b). Following the global scaling of the Hamiltonian as in Eq.~\eqref{eq:scale_H}, $\tH \sim \mathcal{O}(1)$ irrespective of $\gamma$ such that $\tR \simeq \tauR \sim \beta \ll 1$ where $\beta$ is the Dyson index. 
Below $\tR$, the SFF exhibits a correlation hole reflecting the presence of long-range correlation. We show that by monitoring the gradual depletion of the correlation hole, we can identify the localization transition at $\gamma = 1$. 
%relative depth and width of the shrink as we approach the localized regime. 
We also look at the 4-point SFF, whose non-equilibrium features and scaling of the characteristic timescales are similar to its two-point counterpart. As we approach the localized regime, the relaxation time shifts away from the Heisenberg time, in stark contrast to the typical many-body systems. The absence of energy correlation below the critical energy $\Ecal_c$ is reflected in the time window $[\tR, \tH]$, which we term the {\itshape \corrvoid}. The anomalous critical energy found in the \bte\ is also expected to be present in other random matrix models like the banded matrices~\cite{Das2025b}, 
%Burin-Maksimov model~\cite{Tang2022}, 
constrained ensembles with correlated entries~\cite{Mondal2017} and physical systems like kicked rotor~\cite{Pandey2017, Kumar2020}, complex atoms~\cite{Chirikov1985}, isolated thick wire~\cite{Iida1990, Fyodorov1993}.

The behavior of the SFF provides experimentally accessible signatures of nonergodicity, offering pathways to probe thermalization in synthetic quantum platforms. As a future direction, we would like to study the finite-temperature SFF in the \bte, with particular emphasis on the effect of temperature on the timescales and equilibrium behavior of the SFF. 
%=============================
%	---ACKNOWLEDGEMENT---
%=============================
\begin{acknowledgements}
	We acknowledge the support from the Kepler Computing facility, maintained by the Department of Physical Sciences, IISER Kolkata, for various computational needs. A.~K.~D. is supported by an INSPIRE Fellowship, DST, India.
    I.~M.~K. acknowledges support by the European Research Council under the European Union’s Seventh Framework Program Synergy ERC-2018-SyG HERO-810451.
\end{acknowledgements}

%======================================
%%%%% APPENDIX %%%%%
%======================================
%\appendix
%======================================
%%%%% REFERENCES %%%%%
%======================================
\bibliography{ref_SFF_beta}

\end{document}